We address problems arising in supersaturated systems of small atomic particles in solids. Condensation processes in such systems do not seem to suffice a classical interpretation and may be indicative of quantal nucleation particularly at higher supersaturations. We reconcile Gibbs' free energy ΔG vs. particle radius r dependence with the double-well oscillator energy vs. configuration coordinate q dependence to take advantage of the solution of a well-known eigenvalue problem. Theoretical results are presented and compared with experimental data.

1. Introduction

Classical nucleation belongs to the multitude of well-developed branches of present day chemical and solid state physics [1]. Its premises have been laid down by both clever experiments and ingenious theories. Yet the behavior of strongly supersaturated systems have not been understood well despite expectations of the predominance of quantal processes adding to the dissatisfaction of an incomplete exploration. To fill in some of this gap in modern knowledge is the main goal of the present investigation.

2. Classic nucleation

We consider the simple model of a lattice gas of atomic particles and their aggregates (clusters), each one in the form of a sphere of radius r and number of atoms $x = r^3/r_0^3$ where $r_0$ is the "atomic radius" from the atomic volume $v_0 = (4/3)\pi r_0^3$. Gibbs' free energy of a cluster is [2]

$$\Delta G(r) = -(4/3)\pi r^3 (q\Delta T/T_0) + 4\pi r^2 \sigma = -(4/3)\pi r^3 s + 4\pi r^2 \sigma \qquad (1)$$

The 1st term in round brackets represents the supersaturation $s \equiv s(q,T) = q\Delta T/T_0$, $\Delta T = T-T_0$ ($q$ is the heat of nucleation, T is the ambient temperature, $T_0$ is the equilibrium temperature).

The dependence $\Delta G(r)$ is shown in Figure 1. We see a multi-extremum curve with minimum at $r_O = 0$ and maximum at

$$r_C = 2\sigma/(q\Delta T/T_0) = 2\sigma / s \qquad (2)$$

The maximum defines a "critical nucleus". From (1) and (2) we also obtain the work for creating a critical nucleus:

$$\Delta G(r_C) = (4/3)\pi\sigma r_C^2 = (16/3)\pi\sigma^3/s^2 \qquad (3)$$

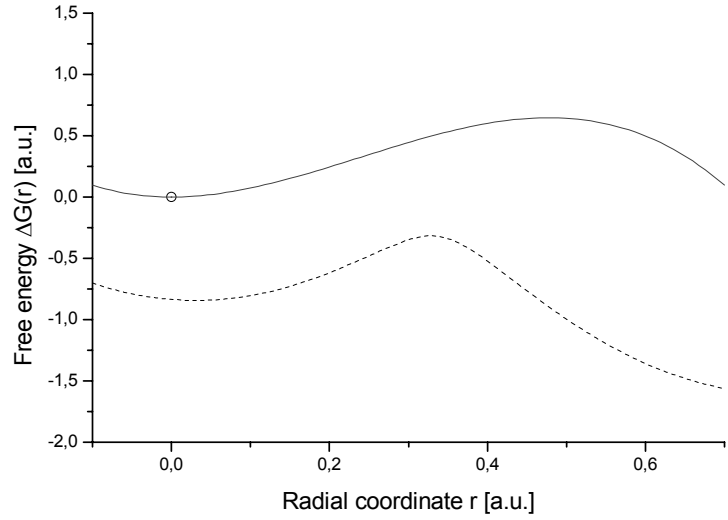

Figure 1: Free energy ΔG vs. radial configuration coordinate r in nucleation (solid curve) and double well (dashed curve) The two dependencies are similar even though not identical to each other. For the drawing, we used K = 17 eV/Å$^2$, G = 6.38 eV/Å, $E_{gap}$ = 0.5 eV, D = 0.77 eV (migration) and σ = 0.676 eV/Å$^2$, s = 2.83 eV/Å$^3$ (nucleation).

It amounts to one-third of the total surface energy of the critical nucleus. Actually, $\Delta G(r_C)$ is the barrier height at r = $r_C$. This barrier is seen to be the lower the higher the supersaturation. On the other hand, tunneling will be favored by lower barriers, as so will the quantal effects.

### 3. Radial nucleation potential in Schrödinger's equation

We may regard equation (1) and the underlying physics as the eigenvalue problem of a radial potential. If so it may be appreciated that the surface term gives rise to a harmonic oscillator, while the bulk term produces an anharmonic vibrator. The eigenstates of the former one are available from textbooks on quantum mechanics [3], while the eigenstates of the latter one may be found as perturbations. Accordingly, we have

$$-(\hbar^2/2M)\nabla^2\psi + \Delta G(r)\psi = \varepsilon\psi \qquad (4)$$

In any event, the radial vibronic potential energy curve in Figure 2(a) is akin to a double well potential of the strongly *exothermic* type (from left to right) with a left-hand well centered at r = 0, an interwell barrier at r = $r_C$ and a right-hand well at r = $r_S$ [4]. The position of the latter well is determined by the nuclei radius about which the super-saturation begins to fall down due to consumption. Further increases of the nuclei radii pushes the potential energy upwards. A new metastable state, therefore, arises at r = $r_S$. The difference $\Delta G(r_O) - \Delta G(r_S) = Q > 0$ is the reaction heat at 0 K. There will be a spontaneous growth of nuclei at Q > 0 until the super-saturation is consumed.

It should be noted that the above configuration is not unique, in as much as depending on the saturation the reaction heat Q may be found negative, $\Delta G(r_O) - \Delta G(r_S) = Q < 0$, if the metastable well at r = $r_S$ happens to be not below $\Delta G(r_O)$ but above it. Now we have an

*endothermic* situation in which the system is under-saturated. In this state at Q < 0 the nuclei can grow only at the expense of an external perturbation. For an illustration, see Figure 2(b).

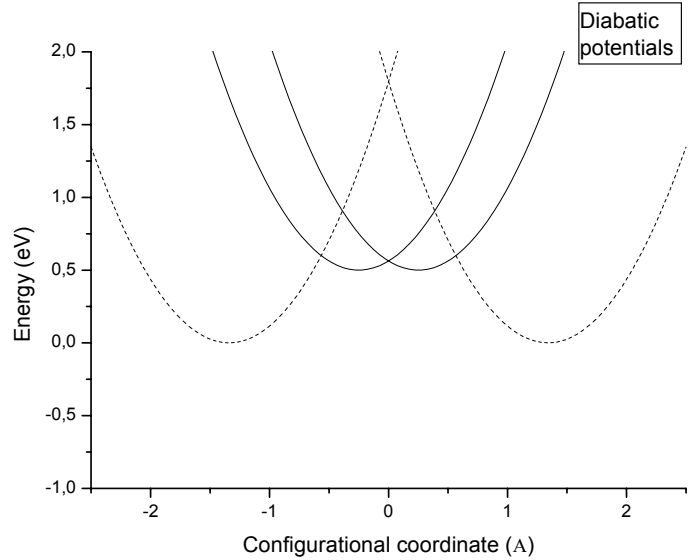

Figure 2: Three configuration situations important for illustrating reactive processes in a solid: left-right dashed and center-center solid are called isothermic, left (dashed)-center (solid) are endothermic, center (solid)-right (dashed) are endothermic.

In the intermediate state at Q = 0 there is an equilibrium between the mother phase and the nuclei of the daughter phase. The process from left to right has been called *isothermic* [4-6]. The intermediate case can be found illustrated in Figure 2(c).

In what follows we shall chiefly explore the approach to exothermic situations, even though extensions to the other two configurations may be achieved quite easily.

Our further considerations will center on the summary transitions from left to right, that is, on the *nucleation rate*, exo-, endo-, or iso- thermic one. In so far as classic rates are available outright, we shall concentrate on the quantal rates instead. The math apparatus we will be using is that of reaction-rate theory.

4. Quantum-mechanical migration / nucleation rate

The reaction-rate theory is an occurrence probability approach aimed at the rate of quasi-chemical reactions in solids [4]. Being based on probabilistics, it is physicslly transparent and is often preferred to other more traditional methods, such as the multiphonon theories (MPT) [4-7]. Most often, the rate theory applies to two-site problems described by double-well potentials of the form compared with equation (1) in Figure 1:

$$E_{\pm}(q) = \tfrac{1}{2} Kq^2 \pm \tfrac{1}{2} \sqrt{[(2Gq+D)^2 + E_{gap\alpha\beta}^2]} \qquad (5)$$

where K is the stiffness of the coupled vibration, G is the coupling constant, D is a constant mixing term, coming from some configurational asymmetry, $E_{gap}$ is an energy gap, q is the configuration coordinate. Equation (5) is the $1^{st}$-order-perturbation ground-state energy – solution to the eigenvalue problem of two nearly-degenerate (finite gap $E_{gap\alpha\beta}$) different-parity fermion states α and β mixed through coupling to an odd-parity boson mode q, the vibronic problem. Traditionally, the fermions are taken to be electrons, while crystalline phonons appear as bosons. Comparing equations (1) to (4) we see a similarity in that r ↔ q, K ↔ 8πσ, mode stiffness, G = $r_C$K ↔ $16\pi\sigma^2/s$, coupling constant, hv ≡ $h\omega = h\sqrt{(8\pi\sigma/M)}$ (M is the component atomic mass), phonon energy, $E_R = Kr_C^2$ ↔ $64\pi\sigma^3/s^2$, lattice reorganization energy.

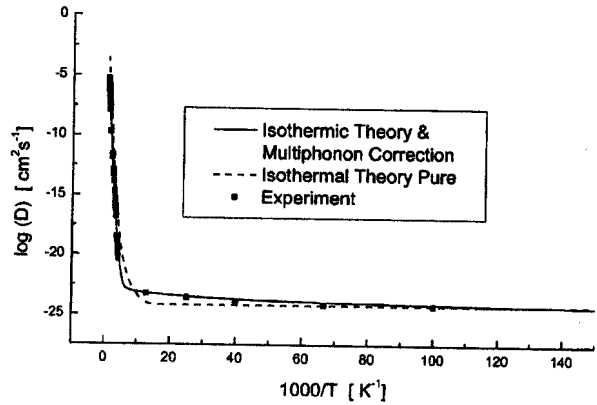

Figure 3: Reaction rate fit to experimental data on the diffusion of carbon in iron: It unravels the participation of 1-phonon and 0-phonon processes below 100 K followed by a steep Ahhrenius branch above 200 K.

In so far as the nonradiative reaction rate $\Re(T)$ and the traditional nucleation rate $\aleph(T)$ apply to similar, if not identical graphic objects, we shall formally make use of $\Re$ to tackle nucleation problems as well, provided the entering reaction parameters are all replaced by their nucleation substitutes, as above.

### 4.1. General migration / nucleation rate

We follow Christov using his reaction rate approach (RRA), while introducing the appropriate changes to make it adaptable to the present problem, to define a general migration rate (potemtially nucleation rate as well):

$$\Re(T) = \nu (Z_A/Z_O) \sum_{n=1}^{\infty} W(E_n)\exp(-E_n/k_BT) \tag{6}$$

where ν is the frequency of the coupled vibrational mode, $(Z_A/Z_O)$ is the fraction of reaction active modes A out of the total modes multitude O in terms of the partition functions $Z_A$ and $Z_O$. For phonon modes $Z_A/Z_O = 2\sinh(\frac{1}{2}hv / k_BT)$. $E_n$ are the quantized eigenenergies of the double-well oscillator.

The eigenvalue spectrum of a harmonic oscillator is well-known, albeit this is not the case for the anharmonic component in equation (1). For the present purposes we may use the harmonic quantities alone and this will pay back, in so far as the approximation is acceptable. In any event, the eigenvalue spectrum of the anharmonic mode shall be dealt with in Appendix I.

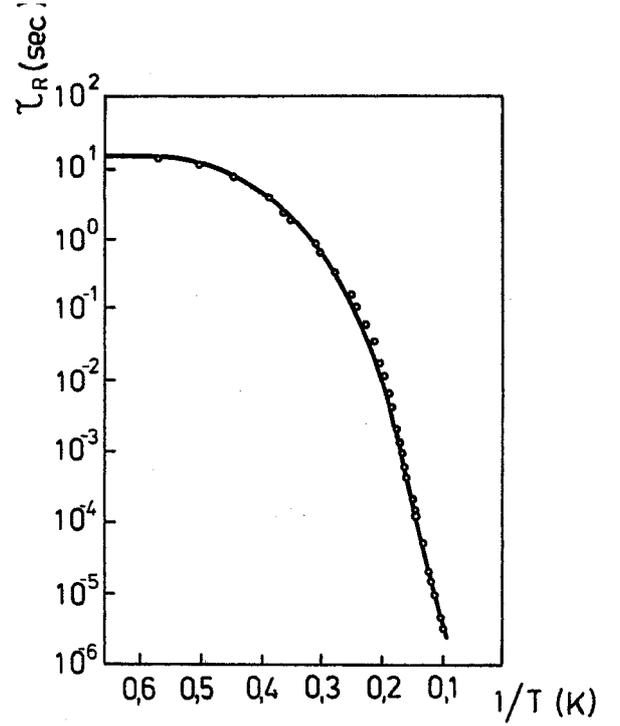

Figure 4: Relaxation time vs. temperature data for off-center $Ag^+$ in RbBr. The 0-phonon contribution below 1 K is evidenced along with the Arrhenius slope above 5 K.

### 4.2. Transition probabilities

We consider only strong-coupling configuration cases where the crossover coordinate is between the two well bottoms, left and right ones. For strong coupling situations we will assume the validity of Condon's approximation which factorizes out the electronic and configurational terms, as in:

$$W(E_n) = W_{el}(E_n)W_L(E_n) \qquad (8)$$

$W_{el}(E_n)$ is the probability (Landau-Zenner's) for a change of the electron state during the left-to-right transition across the barrier in Figures 2(a) through (c), $W_L(E_n)$ is the probability for configurational (lattice) tunneling across that same barrier.

$$\gamma(E_n) = (V_{12}/2h\nu) \sqrt{\{1/E_R|E_n-E_C|\}}$$

is Landau-Zenner's parameter. It has been obtained:

for overbarrier transitions at $E_n \gg E_C$:

$$W_{el}(E_n) = 2[1 - \exp\text{-}(2\pi\gamma)] / [2 - \exp\text{-}(2\pi\gamma)], \qquad (9)$$

for subbarrier transitions at $E_n \ll E_C$:

$$W_{el}(E_n) = 2\pi\gamma^{2\gamma-1} \exp\text{-}(2\gamma) / \gamma\Gamma(\gamma)^2$$

$$W_L(E_n) = \pi \{F_{nm}(\xi_{f0}, \xi_C)^2 / 2^{n+m} n! m!\} \exp(-[n-m]^2 h\nu/E_R)\exp(-E_R/h\nu) \quad (10)$$

Here as an energy conservation condition $Q = (n - m)h\nu$ stands. n and m are the quantum numbers in the initial and final electron states, respectively. Further on (see Appendix II)

$$F_{nm}(\xi_{f0}, \xi_C) = \xi_0 H_n(\xi_C)H_m(\xi_C-\xi_0) - 2nH_{n-1}(\xi_C)H_{m-1}(\xi_C-\xi_0) + 2mH_n(\xi_C)H_m(\xi_C-\xi_0) \quad (11)$$

where $\xi = \sqrt{(M\omega^2/h\nu)}q$ is the dimensionless phonon coordinate, $\xi_{i0} = 0$ and $\xi_{f0} = \xi_0$ are the well-bottom phonon coordinates in the initial and final electronic states, respectively. Here and above $H_i(\xi)$ are Hermite polynomials of n-th order. The above equations hold good at $V_{12} \ll E_C = \Delta G(r_C) - \Delta G(r_O)$. This condition leaves blank a considerable portion of the energy axis.

## 5. Alternative rate equations

The above reaction-rate equation is not free of deficiency having a limited application at energies sufficiently far from the barrier top only, both below and above it: $|E_n-E_C| \gg h\nu$. For this reason it may not be expected to work well in the vicinity of the top, by far the most important energy range in a realistic situation. "*If you have no other means to tackle a rate then use quasiclassics, it's better than nothing*", says a popular advise to the theorist. The basic quasiclassic equations are reproduced in Appendix III. Being somewhat less accurate in the vibronic ground state, the QC predictions are preferable for the vibronic excited states, while the QM predictions are always more reliable in vibronic ground state, far below the barrier top.

Alternative approaches have also been used within the reaction-rate framework [8]. Lately, we proposed an extension to combine the horizontal transitions rate as in equation (6) with the multiphonon vertical-tunneling theory [9]. We found a satisfactory agreement as shown in Figure 3 for the migration of carbon impurity in α-iron [10]. Some other examples can be found elsewhere [11].

## 6. Nucleation vs. migration

The iron work is worth mentioning for it raises an important question as to whether the RRA can distinguish between migration and nucleation without any additional information. Indeed as far as the tunneling barrier is concerned, it may be regarded as either a migrational barrier or the work done for creating the critical cluster in nucleation. We will attempt to reformulate the theory so as to comprise either migration or nucleation and find a way to distinguishing between them, apart from providing an additional information to tell which measured barrier meets which one of the two criteria.

The improved RRA rate accounting for both horizontal- and vertical- tunneling processes fits rather well the experimental carbon-in-iron data, as shown in Figure 3. It will be seen that as the temperature is raised from very low values there is an initial slow rise in rate up to about 100 K and following a transition region in which the rate grows more rapidly a steep growth settles down in which the rate is exponential with the reciprocal temperature. This is the range where the observed slope is proportional to either the migration barrier $E_B$ or the nucleation work $\Delta G(r_C)$. Aside from the transition region which is harder to tackle, the missing data to reveal the origin of the slope observed may be extracted from either the lowest-temperature rate combined with the highest temperature Arrhenius branch. The zero-point tunneling rate is

$$\Re(0) = (E_R/h) \exp(-E_R/h\nu). \tag{12}$$

at Q = 0 where $E_R = M\omega^2(q_C)^2$ in migration interpretation is the lattice reorganization (relaxation) energy. This is about the energy expended for creating the lateral well-bottom site away from the crossover point at $q_C$. Its value has been found in nucleation terms as well to be $E_R = 64\pi\sigma^3/s^2$ from $h\nu \equiv h\omega \leftrightarrow h\sqrt{(8\pi\sigma/M)}$ and $q_C \leftrightarrow 2\sigma/s$. Inserting into (12) we next obtain the zero-point nucleation rate through tunneling (isothermic run):

$$\aleph(0) = [32\pi\sigma^3/s^2]/h \times \exp(-[32\pi\sigma^3/s^2]/h\sqrt{[8\pi\sigma/M]}). \tag{13}$$

Table 1

Inter-conversion between migration and nucleation quantities

| Quantity | Migration | Nucleation |
|---|---|---|
| Phonon energy | $h\nu \equiv h\omega$ (meV) | $h\sqrt{(8\pi\sigma/M)}$ |
| Stiffness | $K = M\omega^2$ (eV/Å²) | $8\pi\sigma$ |
| Coupling constant | G (eV/Å) | $16\pi\sigma^2/s$ |
| Crossover barrier | $E_B$ (eV) | $\Delta G(r_C) - \Delta G(r_O)$ |
| Crossover energy | $E_C = E_B + \frac{1}{2}E_{gap}$ (eV) | - |
| Barrier top coordinate | $q_C$ (Å) | $2\sigma/s$ |
| Reorganization energy | $E_R$ (eV) | $32\pi\sigma^3/s^2$ |
| Zero-point reaction heat | Q (eV) | $\Delta G(r_O) - \Delta G(r_S) =$ |
| Zero-point rate | $\Re(0) = (E_R/h) \exp(-E_R/h\nu)$ | $\aleph(0) = [32\pi\sigma^3/s^2]/h \times \exp(-[32\pi\sigma^3/s^2]/h\sqrt{[8\pi\sigma/M]})$ |

σ- specific surface energy, q- specific nucleation heat, T- absolute temperature, $s = q\Delta T/T_0$- supersaturation.

7. Carbon migration/nucleation in iron

The temperature dependence of the reaction rate is displayed in Figure 3. Our interpretation will be based on the tabulated formulae in Table 1 and data in Table 2. The latter were obtained by fitting equation (6) to the experimental temperature dependence.

Table 2

Numerical data for RRA parameters as obtained from diffusion experiment

| Quantity | Migration | Nucleation |
|---|---|---|
| Photon quantum | $h\nu = 77$ (meV) | $h\sqrt{(8\pi\sigma/M)} \to$ $M = (8\pi\sigma)/\omega^2$ |
| Stiffness | $K = 17$ eV/Å$^2$ | Surface energy $\sigma = 0.676$ (eV/Å$^2$) |
| Coupling constant | $G = 6.272$ eV/Å | Supersaturation $s = q\Delta T/T_0 = 1.83$ eV/Å$^3$ |
| Crossover barrier | $E_B = 1.1$ eV | $\Delta G(r_C) = (4/3)\pi\sigma r_C^2$ $\Delta G(r_C) = 1.54$ eV |
| Crossover energy | $E_C = 1.158$ eV | Energy gap $E_{gap} = 2(E_C - E_B)$ 116 (meV) |
| Barrier top coordinate | $q_C = 0.738$ Å $(q_O = 0)$ | $r_C = q_C$ |
| Reorganization energy | $E_R = \frac{1}{2}Kq_C^2$ $= 4.629$ eV | $32\pi\sigma^3/s^2$ $= 2.57$ meV |
| Zero-point reaction heat | $Q = 0$ | $Q > 0$ |
| Zero-point rate | $\mathfrak{R}(0) = 9.5 \times 10^{-9}$ s$^{-1}$ $\mathfrak{R}(0) = 5.86 \times 10^{-11}$ s$^{-1}$ | $\aleph(0) = 2 \times 28.819$ s$^{-1}$ $\aleph(0) / \mathfrak{R}(0) = 6 \times 10^9$ |

The zero-point data yield $\aleph(0) / \mathfrak{R}(0) = 6.06 \times 10^9$ which looks like a scale factor originating from the discrepancy between diffusion and nucleation data. The discrepancy between quantities in the two sets of theorized data may be indicative of the inconsistent alternative interpretation through nucleation within the *whole range* of parameters. Migration may be convincing in cases where nucleation fails, and vice versa. We believe the former may apply to cases with a strong 1-phonon contribution leading to a slope, though minor, within the low temperature tunneling range. If 1-phonon contribution helps nucleation, as may be expected, then we will have a strong motif for future studies. Alternatively, a vanishing slope may give evidence for a sole horizontal-tunneling behavior. In other cases both appearances may be predominating in different temperature ranges. All this makes the reaction rate propositions so thrilling and candidates for interesting further developments. In Figure 4 we show the case of a flat zero-point rate, as obtained earlier [11].

8. Conclusion

The possibility for defining a nucleation rate by means of a reaction rate formula is a sole advantage of the probabilistic nature of the latter. Indeed, the problem is one of calculating two physical events, i.e. tunneling in migration and nucleation, across virtually the same potential energy surface which events are controlled by the transition probabilities across that surface. From this point of view it does not matter a thing what meaning you may give to the process if its essential part is the same in both cases: configurational tunneling across a pre-given barrier surface. Indeed, the tunneling probabilities are independent of the nature of the process, while the controlling parameters are defined and derived by the most general features of the surface: positions of barrier and metastable valleys mainly, as well as the dependence on the acting coordinate, close to a double-well parabola.

In short, vibronic surface and related configuration tunneling probabilities $W_{conf}(E_n)$ go hand in hand whatever meaning may be given to the processes attached to them. This refers chiefly to the rates of adiabatic processes with ultimate electron transfer probabilities, $W_{el}(E_n) = 1$.

In most of the processes associated with a double-well potential we studied nonradiative deexcitation rates but also rotational tunneling rates, as a generalization of the two-site tunneling problem [11]. In that we also introduced and studied Mathieu-function based rotational rates of off-center species [12]. We found a close agreement between data obtained by different tunneling means suggesting converging results based on Bardeen's method [4]. This may also give credit to our present attempt to extending this method to nucleation or aggregation rates in solids as well.

## Appendix I

## Eigenvalue spectrum for the anharmonic $q^3$ mode

To first order, the 'A' eigenvalues may be found as

$$E_\pm = \tfrac{1}{2}(H_{11} + H_{22}) \pm \tfrac{1}{2}\sqrt{[(H_{11} - H_{22})^2 + 4H_{12}H_{21}]} \qquad (AI.1)$$

where $|1\rangle$ and $|2\rangle$ are appropriate basis states, e.g. displaced harmonic-oscillator eigenstates. In this case the matrix elements $H_{ij}$ read $H_{11} \equiv H_{++}$, $H_{22} \equiv H_{--}$, $H_{12} = H_{21} = H_\pm$, to small-polaron approximation. We introduce the harmonic-oscillator ground states and the full Hamiltonian:

$$\chi_\pm(q\pm q_0) = A\exp(-\tfrac{1}{2}\alpha[q \pm q_0]^2) \qquad (AI.2)$$

$$H = (-\hbar^2/2M)\nabla^2 \chi(q) + [Bq^2 - Cq^3]\chi(q) = (n+\tfrac{1}{2})h\nu\,\chi(q) \qquad (AI.3)$$

forming the following matrix elements (small-polaron approximation)

$$H_{++} = \langle 1|H|1\rangle = \tfrac{1}{2}h\nu + A^2\sqrt{(1/\alpha)}\exp(-\alpha q_0^2)\int dq\,q^3\exp(-q^2) \approx \tfrac{1}{2}h\nu + A^2\sqrt{(\pi/\alpha)}\exp(-\alpha q_0^2)q_0^3$$

$$H_{--} = \langle 2|H|2\rangle = \tfrac{1}{2}h\nu + A^2\sqrt{(\pi/\alpha)}\exp(-\alpha q_0^2)\int dq\,q^3\exp(-q^2) \approx \tfrac{1}{2}h\nu + A^2\sqrt{(\pi/\alpha)}\exp(-\alpha q_0^2)q_0^3$$

$$H_\pm = \langle 2|H|1\rangle = \langle 1|H|2\rangle \approx [\tfrac{1}{2}h\nu + A^2\exp(-\alpha q_0^2)\sqrt{(\pi/\alpha)}]\,S_{12}\int dq\,q^3\exp(-q^2) \qquad (AI.4)$$

where $S_{12}$ is the overlap integral. Accordingly we get

$$E_\pm = \tfrac{1}{2}h\nu + A^2\sqrt{(\pi/\alpha)}\exp(-\alpha q_0^2)q_0^3 \pm |\tfrac{1}{2}h\nu + A^2\sqrt{(\pi/\alpha)}\exp(-\alpha q_0^2)q_0^3|$$

$$= |h\nu + 2A^2\sqrt{(\pi/\alpha)}\exp(-\alpha q_0^2)q_0^3|\times\binom{1}{0} \qquad (AI.5)$$

Here and above $A_n = [\sqrt{(\alpha/\pi)}(1/2^n n!)]^{1/2}$ gives the normalization constant $A_0$ at n=0.

## Appendix II

### Transition probabilities

Generally, the transition probability $W(E_n)$ is defined in terms of the flux of vibrons in the initial electron state (left) along the reaction coordinate q towards the transition configuration at $q_C$. This flux is partially reflected back from the barrier and partially transmitted to the final electron state (right). The reverse current back from the final state may be neglected, if assumed that once in that state the vibron relaxes rapidly to lower levels giving away the excess energy through its coupling to the accepting modes, so that the chances for return are rather small. Under these conditions, the tunneling probability reads

$W(E_n) = j_{transmitted} / j_{incident}$

$j(q) = \frac{1}{2} i \sqrt{(h\nu/M)} [\chi d\chi^*/dq - \chi^* d\chi/dq]$ (AII.1)

Undoubtedly, all the underlying quantities can be found by solving Schrödinger's equation with radial potential (4). We presented the final results in 4.2 though the arguments involved are given in this Appendix II. They all follow suggestions originating from John Bardeen [4].

We have seen that the motion along the configuration coordinate $q = r$ is barrier controlled. In particular, the configurational transition probability along the radial coordinate based on the currents across the barrier will be [5]

$W_{if\,conf}(E_n) = 4\pi^2 |V_{fi}|^2 \sigma_i(E_n)\sigma_f(E_n)$ (AII.2)

where the matrix element $V_{fi}$ is to be calculated using initial and final state wave functions $\phi_i$ and $\phi_f$, respectively, as:

$V_{fi} = (-h^2/2I) [\phi_f^* (d\phi_i/dq) - \phi_i (d\phi_f/dq)^*]|_{q=qc}$ (AII.3)

Here $\sigma_i$ and $\sigma_f$ are the corresponding density-of-states (DOS) of the initial and final states. For a harmonic oscillator $\sigma_i(E_n) = \sigma_f(E_n) = (h\nu)^{-1}$. Inserting into (AII.1) and performing the math in (AII.2) we obtain the formulas (10) – (11) of the text.

## Appendix III

### Quasi-classical rates

Herein we briefly revisit the basic QC equations. These reflect changes in philosophy of the approach to the configuration probabilities. We define a configurational probability term

$W_{conf}(E_n) = 1/\{1 + \exp(K_1 + K_2)\}$ (AIII.1)

where $K_i(E_n)$ are the phase integrals

$K_i(E_n) = \sqrt{(2M/h^2)} \int_{qi(En)}^{qC} \sqrt{[E_\pm(q) - E_n]}\, dq$ (AIII.2)

$q_1$ and $q_2$ are the classical turning points, $q_C$ is the crossover point (see a drawing in Figure 5).

Performing the integration in (AIII.2) we get

$$K_1(E_n) = (2E_C/h\nu)\{\sqrt{\alpha_1} - (1 - \alpha_1) \ln [\sqrt{(1 - \alpha_1)} / (1 - \sqrt{\alpha_1})]\}$$

$$K_2(E_n) = [2(E_C - Q)/h\nu]\{\sqrt{\alpha_2} - (1 - \alpha_2) \ln [\sqrt{(1 - \alpha_2)} / (1 - \sqrt{\alpha_2})]\}$$

(AIII.3)

where $\alpha_1 = (E_C - E_n) / E_C$, $\alpha_2 = (E_C - E_n) / (E_C - Q)$.

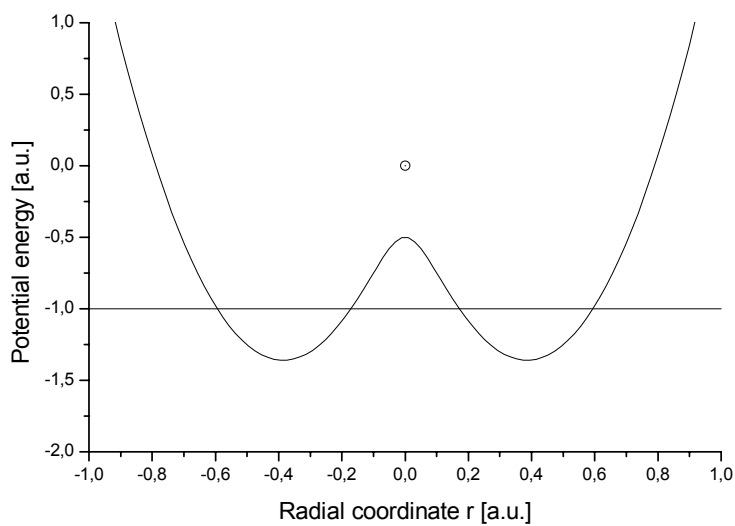

Figure 5: Arrangements for a quasi-classic calculation: crossover points from left to right – left-hand classic turning point, energy level on the left-hand slope of barrier, energy level on the right-hand slope of barrier, right-hand classic turning point.